\documentclass[a4paper,11pt]{article}

\usepackage{jheppub}
\usepackage[table]{xcolor}
\usepackage{slashbox}
\allowdisplaybreaks
\usepackage{graphicx}
\usepackage{epstopdf}

\newtheorem{definition}{Definition}

\newcommand{\be}{\begin{equation}}
\newcommand{\ee}{\end{equation}}
\newcommand{\bea}{\begin{eqnarray}}
\newcommand{\eea}{\end{eqnarray}}
\newcommand{\nn}{\nonumber}
\newcommand{\bdm}{\begin{displaymath}}
\newcommand{\edm}{\end{displaymath}}

\title{N$^3$LO gravitational spin-orbit coupling at order $G^4$} 
\author[a,b]{Mich\`ele Levi,}
\author[a]{Andrew J.~McLeod,}
\author[a]{and Matthew von Hippel}

\affiliation[a]{Niels Bohr International Academy, Niels Bohr Institute,
University of Copenhagen,
\\Blegdamsvej 17, 2100 Copenhagen, Denmark}

\affiliation[b]{Institut de Physique Th\'eorique, 
CEA \& CNRS, Universit\'e Paris-Saclay,
\\ 91191 Gif-sur-Yvette, France}

\emailAdd{michelelevi@nbi.ku.dk}
\emailAdd{amcleod@nbi.ku.dk}
\emailAdd{mvonhippel@nbi.ku.dk}

\abstract{
In this paper we derive for the first time the N$^3$LO gravitational spin-orbit coupling at 
order $G^4$ in the post-Newtonian (PN) approximation within the effective field theory (EFT) 
of gravitating spinning objects.
This represents the first computation in a spinning sector involving three-loop integration.
We provide a comprehensive account of the topologies in the worldline picture for the 
computation at order $G^4$.
Our computation makes use of the publicly-available \texttt{EFTofPNG} code, which is extended 
using loop-integration techniques from particle amplitudes. 
We provide the results for each of the Feynman diagrams in this sector. 
The three-loop graphs in the worldline picture give rise to new features in the spinning 
sector, including divergent terms and logarithms from dimensional regularization, as well as 
transcendental numbers, all of which survive in the final result of the topologies at this 
order. %We also find that this piece of the sector vanishes in the test-particle limit.
This result enters at the 4.5PN order for maximally-rotating compact objects, and together 
with previous work in this line, paves the way for the completion of this PN accuracy.} 

\begin{document}

\today

\maketitle

\flushbottom

\section{Introduction} 
\label{intro}

The recent detection of gravitational waves (GWs) by the LIGO collaboration 
\cite{Ligo,Abbott:2016blz} has reinforced the importance and urgency of high-precision 
gravity predictions. The worldwide network of ground-based detectors has been continually 
growing \cite{Virgo,Kagra} with upcoming ground- and space-based detectors to reach a broader 
band of frequencies, as well as higher sensitivities 
\cite{IndiGO,Punturo:2010zz,PAIK:2016yxn,Lisa,Luo:2015ght}. 
These observations are primarily designed to detect the mergers of binaries of compact 
components, which build up to this dramatic event via a long inspiral phase in which they 
orbit each other with non-relativistic velocities. For this reason the analytical study of 
the post-Newtonian (PN) approximation of General Relativity has become crucial, including the 
orbital dynamics of the compact binaries \cite{Blanchet:2013haa}, an essential ingredient for 
the theoretical waveform models that are created using the effective one-body (EOB) framework 
\cite{Buonanno:1998gg}.

\begin{table}[t]
\begin{center}
\begin{tabular}{|l|r|r|r|r|r|r}
\hline
\backslashbox{\quad\boldmath{$l$}}{\boldmath{$n$}} &  (N\boldmath{$^{0}$})LO
& N\boldmath{$^{(1)}$}LO & \boldmath{N$^2$LO}
& \boldmath{N$^3$LO} & \boldmath{N$^4$LO} & \boldmath{N$^5$LO} 
\\
\hline
\boldmath{S$^0$} & 1 & 0  & 3 & 0 & 25 & 0
\\
\hline
\boldmath{S$^1$} & 2 & 7 & 32 & 174 & & 
\\
\hline
\boldmath{S$^2$} & 2 & 2 & \textbf{18} & \textbf{52} & &
\\
\hline
\boldmath{S$^3$} & 4 & \cellcolor[gray]{0.9} \textbf{24} & \cellcolor[gray]{0.9} 
& \cellcolor[gray]{0.9} & \cellcolor[gray]{0.9} & \cellcolor[gray]{0.9}
\\
\hline
\boldmath{S$^4$} & 3 & \textbf{5} \cellcolor[gray]{0.9} & \cellcolor[gray]{0.9} 
& \cellcolor[gray]{0.9} & \cellcolor[gray]{0.9} & \cellcolor[gray]{0.9} \\
\end{tabular}
\caption{The number of highest-loop graphs in the worldline picture contributing 
to each sector in the state-of-the-art of PN gravity for the orbital dynamics of 
compact binaries. 
}
\label{stateoftheart}
\end{center}
\end{table}

Table \ref{stateoftheart} shows the complete state of the art in PN orbital dynamics of 
generic compact binaries to date. Each correction enters at the $n+l+\text{Parity}(l)/2$-th 
PN order, where $n$ indicates the N$^n$LO, $l$ indicates the order in spin up to the $l$-th 
multipole $S^l$, and the parity is $0$ or $1$ for even or odd $l$, respectively. 
The table lists the number of the highest-loop graphs (as defined in section \ref{topologies} 
below) in each sector as a measure of its computational scale in the effective field theory 
(EFT) framework for PN gravity \cite{Goldberger:2004jt,Levi:2018nxp}, with $n$ loops entering 
generically at N$^n$LO. Note that, in the non-spinning sector, when the Kaluza-Klein (KK) 
field decomposition \cite{Kol:2007bc,Kol:2010ze} is employed, only the $2\lfloor n/2 
\rfloor$-loop level is required, such that e.g., at the 5PN order only the four-loop level is 
relevant, and so the entry on the first row at N$^5$LO is $0$ \cite{Levi:2018nxp}. 
All sectors up to the 4PN order (apart from the top right one in the non-spinning sector) are 
available within the public \texttt{EFTofPNG} code \cite{Levi:2017kzq,Levi:2018stw}. 

So far the sectors with boldface entries in table \ref{stateoftheart} have been completed for 
generic compact binaries only via the EFT formulation of gravitating spinning objects 
introduced in \cite{Levi:2015msa}, see 
\cite{Levi:2014gsa,Levi:2015ixa,Levi:2016ofk,Levi:2019kgk,Levi:2020uwu,Levi:2020lfn}. 
The formulation in \cite{Levi:2015msa}, which also provided the leading gravitational 
couplings to all orders in spin, thus enabled the completion of the current state of the art 
to the 4PN order. The work in \cite{Levi:2015msa} was also recently extended to the NLO of 
the cubic- and quartic-in-spin sector at the 4.5 and 5PN orders 
\cite{Levi:2019kgk,Levi:2020lfn}, which are to date the only PN works to explore the gray 
area in table \ref{stateoftheart}. The latter is associated with the gravitational Compton 
scattering with spins $s \ge 3/2$, as classical effects with spin to the $l$-th order 
correspond to amplitudes involving a quantum spin of $s=l/2$ \cite{Arkani-Hamed:2017jhn}.

In this work, we derive for the first time via the EFT of spinning gravitating objects the 
N$^3$LO spin-orbit coupling from interaction at $G^4$, which consists of the highest-loop 
graphs in this sector at three-loop level.
This is the highest loop level tackled in the spinning sector so far. 
This coupling enters at the $4.5$PN order for maximally-rotating compact objects, and 
together with the sector tackled in \cite{Levi:2019kgk}, this sector completes the accuracy 
at 4.5PN order, thus uniquely pushing the precision frontier of PN gravity via the EFT of 
spinning gravitating objects. 
This work follows the EFT approach from \cite{Goldberger:2004jt}, building on the EFT of 
gravitating spinning objects \cite{Levi:2015msa} and its implementation in the unique public 
\texttt{EFTofPNG} code in \cite{Levi:2017kzq,Levi:2018stw}, to exploit more methods from 
particle amplitudes. This work also builds on the implementations of \cite{Levi:2015msa} at 
the two-loop level in \cite{Levi:2011eq,Levi:2014sba,Levi:2015uxa,Levi:2015ixa,Levi:2016ofk}, 
and on prior work at the N$^3$LO in the non-spinning sector in
\cite{Jaranowski:1997ky,Jaranowski:1999ye,Blanchet:2000nv,Blanchet:2000ub,Damour:2001bu,
Itoh:2003fy,Blanchet:2003gy,Levi:2011up,Foffa:2011ub}. Possible tails of radiation reaction 
involving spin couplings enter at the 5.5PN order \cite{Blanchet:2011zv,Blanchet:2013haa}, 
namely beyond the order considered in the current sector, so no IR divergences are involved 
in this work.

Beyond the conceptual difficulty of tackling spins in gravity, the spinning sectors are also 
rather challenging at the computational level. It was already highlighted in 
\cite{Levi:2019kgk} that sectors that have an even order in the spin (in particular also the 
non-spinning sector) are consistently easier to handle than those which are odd. First, at 
this loop order it is known that simple poles and their accompanying logarithms in 
dimensional regularization arise at $G^4$ also with traditional PN methods 
\cite{Blanchet:2013haa}. Yet, in the non-spinning sector in an EFT computation with the KK 
decomposition of the metric, such intricate features do not show up at $G^4$ 
\cite{Levi:2011up,Foffa:2011ub}, since as was noted above the KK decomposition postpones the 
appearance of three-loop integration to the N$^4$LO. Thus, while there are no graphs that 
enter at three-loop level at the N$^3$LO without spins, in the spin-orbit sector there are 
$174$ such graphs, as shown in table \ref{stateoftheart}. 

This sector is also more complex in terms of raw number of graphs with $388$ graphs to 
evaluate, compared to only $8$ in the N$^3$LO non-spinning sector. This is because in 
contrast to the non-spinning case, in the spinning sectors all possible topologies are 
realized at each order of $G$ \cite{Levi:2008nh}. Second, since spins are derivatively 
coupled, we tackle integrand tensor numerators as high as rank eight, comparable to N$^5$LO 
in the non-spinning sector. Further, this derivative coupling introduces more time 
derivatives due to the spin couplings. Finally, and related to the previous point, another 
notable aspect that further demonstrates the relative intricacy of the spinning sectors with 
respect to the non-spinning ones is that accelerations enter already at the LO spin-orbit 
sector \cite{Levi:2010zu}, that is at the 1.5PN order, compared to the N$^2$LO in the 
non-spinning sector which is at the 2PN order.

This paper is organized as follows. We begin in section \ref{theory} by presenting the formal 
setup within the EFT of gravitating spinning objects. We then proceed in section 
\ref{Feynman} to study the diagrammatic expansion, first looking into the new topologies in 
the present spinning sector by revisiting the structure of topologies in the EFT approach 
from the ground up in section \ref{topologies}. We go on to consider the specific graphs that 
need to be evaluated in this sector in section \ref{graphs}, and evaluate them by 
supplementing the \texttt{EFTofPNG} code with common methods from particle amplitudes, in 
section \ref{scalability}. We then discuss the findings of our evaluations, highlighting some 
special features in section \ref{findings}, and discuss the total outcome for the sector in 
section \ref{result}. Finally, we conclude in section \ref{lafin}.

\section{EFT of gravitating spinning objects}
\label{theory}

We first present the formal setup required to carry out the EFT computation of the N$^3$LO 
spin-orbit sector at order $G^4$. Here we will build on the presentation in 
\cite{Levi:2015uxa} and \cite{Levi:2018nxp} (as well as the references therein), reviewing 
the relevant part of the one-particle effective action, and introducing the new Feynman rules 
that enter at this order. Note, though, that we will keep here all dependence on the number 
of spatial dimensions, $d$, explicit, as done in the \texttt{EFTofPNG} code 
\cite{Levi:2017kzq}, due to our use of dimensional regularization (with the appearance of 
related divergences, as shall be seen in section \ref{Feynman}).

Let us recall the two-particle effective action describing a compact binary system 
\cite{Goldberger:2004jt,Levi:2018nxp}, which reads
\be \label{2ptact}
S_{\text{eff}}=S_{\text{g}}[g_{\mu\nu}]+\sum_{a=1}^{2}S_{\text{pp}}(\lambda_a),
\ee
where $S_{\text{g}}$ is the pure gravitational action, and $S_{\text{pp}}$ is 
the worldline point-particle action for each of the two components of the binary, which 
depends on a worldline parameter of the $a$-th component, $\lambda_a$. 

First, we need to take into account the purely gravitational action at the 
orbital scale. This is given in terms of the gravitational field, 
$g_{\mu\nu}(x)$, as follows:
\be \label{Sg}
S_{\text{g}}[g_{\mu\nu}]=S_{\text{EH}} + S_{\text{GF}} 
= -\frac{1}{16\pi G_d} \int d^{d+1}x \sqrt{g} \,R 
+ \frac{1}{32\pi G_d} \int d^{d+1}x\sqrt{g}
\,g_{\mu\nu}\Gamma^\mu\Gamma^\nu, 
\ee
where $\Gamma^\mu\equiv\Gamma^\mu_{\rho\sigma}g^{\rho\sigma}$, and we have the 
Einstein-Hilbert action supplemented by a gauge-fixing term, chosen to be 
the fully harmonic gauge. Note that, similar to the modified minimal subtraction 
($\overline{\text{MS}}$) prescription \cite{Peskin:1995ev}, we use here the generic 
$d$-dimensional gravitational constant, $G_d$, defined as:
\be
G_d\equiv G_N \left(\sqrt{4\pi e^\gamma} \,R_0 \right)^{d-3},
\ee
where $G_N\equiv G$ is Newton's gravitational constant in three-dimensional space, 
$\gamma$ 
%\equiv \lim_{z\to 0$}\tfrac{1}{z}-\Gamma(z)$ 
is Euler's constant, and 
$R_0$ is a fixed renormalization scale. In what follows, 
we will use $\epsilon\equiv d-3$ to denote the dimensional parameter in 
dimensional regularization.

We then decompose the gravitational field into a $d+1$ non-relativistic form in 
a Kaluza-Klein (KK) fashion: 
\begin{align}\label{kk}
ds^2=g_{\mu\nu}dx^{\mu}dx^{\nu}\equiv 
e^{2\phi}\left(dt-A_idx^i\right)^2-e^{-\frac{2}{d-2}\phi}\gamma_{ij}dx^idx^j.
\end{align}
This parametrization has considerably facilitated 
higher-order PN computations in the EFT approach \cite{Kol:2007bc,Kol:2010ze}.
Eq.~\eqref{kk} defines the KK fields: $\phi$, $A_i$, and 
$\gamma_{ij}\equiv\delta_{ij}+\sigma_{ij}$, identified as the Newtonian scalar, the 
gravito-magnetic vector, and the symmetric tensor, respectively. The gravitational 
action in eq.~\eqref{Sg} is worked out in terms of the KK fields in \cite{Kol:2010si}, 
and in more detail in the \texttt{EFTofPNG} code \cite{Levi:2017kzq}. 
This action gives rise to the following propagators for the KK fields:
\begin{align}
\label{eq:prphi} \langle{~\phi(x_1)}~~{\phi(x_2)~}\rangle
& = \parbox{18mm}{\includegraphics[scale=0.5]{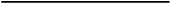}} = 
\quad\frac{16\pi\,G_d}{c_d} \cdot\,\delta(t_1-t_2) \int_{\vec{k}} 
\frac{e^{i \vec{k}\cdot\left(\vec{x}_1 - \vec{x}_2\right)}}{{\vec{k}}^2},\\ 
\label{eq:prA} \langle{A_i(x_1)}~{A_j(x_2)}\rangle 
& = \parbox{18mm}{\includegraphics[scale=0.5]{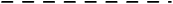}} = 
\,-16\pi\,G_d \cdot \,\delta(t_1-t_2)\int_{\vec{k}} 
\frac{e^{i\vec{k}\cdot\left(\vec{x}_1 - \vec{x}_2\right)}}{{\vec{k}}^2} 
~\delta_{ij},\\ 
\label{eq:prsigma}  \langle{\sigma_{ij}(x_1)}{\sigma_{kl}(x_2)}\rangle 
& = \parbox{18mm}{\includegraphics[scale=0.5]{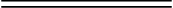}} = 
\quad32\pi\,G_d \cdot\,\delta(t_1-t_2)\int_{\vec{k}}
\frac{e^{i\vec{k}\cdot\left(\vec{x}_1 - \vec{x}_2\right)}}{{\vec{k}}^2} 
~P_{ij;kl},
\end{align}
where we abbreviate $\int \frac{d^d\vec{k}}{\left(2\pi\right)^d}$ as 
$\int_{\vec{k}}$, $P_{ij;kl}\equiv\frac{1}{2}\big(\delta_{ik}\delta_{jl}
+\delta_{il}\delta_{jk}+(2-c_d)\delta_{ij}\delta_{kl}\big)$, and $c_d\equiv 2(d-1)/(d-2)$. 
These propagators receive perturbative relativistic corrections from quadratic vertices 
involving two time derivatives; however, since we only consider the leading contribution in 
$G^4$ to the N$^3$LO sector in this paper such corrections will not play a role. 

Let us turn now to the Feynman rules required for this sector, which go beyond those 
appearing in the lower-order spinning sectors, presented in 
\cite{Levi:2015uxa}. The latter can be found for generic $d$ in the public \texttt{EFTofPNG} 
code \cite{Levi:2017kzq}. The Feynman rules relevant as of this order were also obtained by 
extending the \texttt{FeynRul} module of the \texttt{EFTofPNG} code \cite{Levi:2017kzq}.

We consider first the bulk vertices. There are two new cubic self-interaction vertices to 
consider, 
\begin{align}
\label{eq:sigma^3} \parbox{18mm}{\includegraphics[scale=0.6]{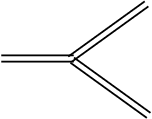}}
 = & \quad  \frac{1}{256\pi G_d}\int d^{d+1}x \Big[
 \sigma_{ii}\big[(\partial_i\sigma_{jj})^2-2(\partial_i\sigma_{jk})^2
 +4(\partial_i\sigma_{ij})^2-4\partial_i\sigma_{jk}\partial_j\sigma_{ik}
 \big]
\nn\\
& \qquad \qquad \qquad 
+2\big[4\sigma_{ij}\partial_k \sigma_{ik}
-4\sigma_{ik}\partial_i \sigma_{jk}-2\sigma_{ik}\partial_j \sigma_{ik}
-\sigma_{ij}\partial_i \sigma_{kk}\big]\partial_j\sigma_{ii}\nn\\
& \qquad \qquad \qquad 
+4\sigma_{ij}\big[\partial_i\sigma_{kl}\partial_j\sigma_{kl} 
-2\partial_k\sigma_{ik}\partial_l\sigma_{jl}
+2\partial_k\sigma_{il}\partial_k\sigma_{jl}
+2\partial_k\sigma_{il}\partial_l\sigma_{jk}
\big]\Big]
,\\
\label{eq:sigma^2A} \parbox{18mm}{\includegraphics[scale=0.6]{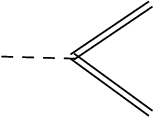}}
 = & \quad \frac{1}{64\pi G_d}\int d^{d+1}x \Big[ \big[ 
  \partial_iA_i\partial_t\sigma_{jj}-\partial_tA_i\partial_i\sigma_{jj}
  +2\partial_tA_i\partial_j\sigma_{ij}-2\partial_iA_j\partial_t\sigma_{ij}
\big]\sigma_{ii}\nn\\
& +A_i \big[\partial_t\sigma_{jj}(\partial_i\sigma_{kk}-2\partial_k\sigma_{ik})
+2\partial_t\sigma_{ij}(\partial_j\sigma_{kk}-2\partial_k\sigma_{jk})
-2\partial_t\sigma_{jk}(\partial_i\sigma_{jk}-2\partial_j\sigma_{ik})\big]
\nn\\
&
+2\partial_tA_i\big[
\sigma_{ij}(\partial_j\sigma_{kk}-2\partial_k\sigma_{jk})
+\sigma_{jk}(\partial_i\sigma_{jk}-2\partial_j\sigma_{ik})\big]\nn\\
& 
-2\partial_jA_i\big[\sigma_{ij}\partial_t\sigma_{kk}
-2\sigma_{ik}\partial_t\sigma_{jk}-2\sigma_{jk}\partial_t\sigma_{ik}\big]
-2\partial_iA_i\sigma_{jk}\partial_t\sigma_{jk}
\Big],
\end{align} 
three new quartic self-interaction vertices, 
\begin{align}
\label{eq:sigmaA^2phi} \parbox{18mm}{\includegraphics[scale=0.6]{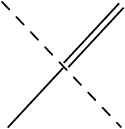}}
 = & \quad \frac{c_d}{64\pi G_d}\int d^{d+1}x 
 \Bigg[\phi \Big[\sigma_{ii} 
 \left(\partial_iA_j\left(\partial_iA_j-\partial_jA_i\right)
  +\left(\partial_iA_i\right)^2\right) \nn\\
  &+2\sigma_{ij}\big[2\left(\partial_i A_k\partial_k A_j
  -\partial_i A_j\partial_k A_k\right)
  -\left(\partial_i A_k\partial_j A_k
    +\partial_k A_i\partial_k A_j\right)\big]\Big]\Bigg]
,\\
\label{eq:sigmaAphi^2} \parbox{18mm}{\includegraphics[scale=0.6]{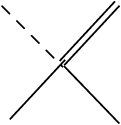}}
  = & \quad \frac{c_d}{32\pi G_d}\int d^{d+1}x 
  \big[\left(2\sigma_{ij}A_i\partial_j\phi
  -\sigma_{ii}A_j\partial_j\phi\right)\partial_t\phi\big],\\
\label{eq:sigma^2phi^2} \parbox{18mm}{\includegraphics[scale=0.6]{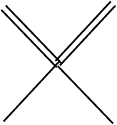}}
  = & \quad \frac{c_d}{256\pi G_d}\int d^{d+1}x 
  \Big[(\partial_i \phi)^2\left(2(\sigma_{ij})^2-(\sigma_{ii})^2\right)
 \nn\\ 
 & \qquad \qquad \qquad \qquad +4\partial_i \phi \partial_j \phi 
  \left(\sigma_{kk}\sigma_{ij}-2\sigma_{ik}\sigma_{jk} \right) \Big],
\end{align}
and one new quintic self-interaction vertex,
\begin{align}
\label{eq:phi^3A^2} \parbox{18mm}{\includegraphics[scale=0.6]{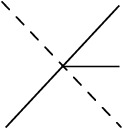}}
 = & \quad \frac{c_d^3}{192\pi G_d}\int d^{d+1}x~\phi^3
 \left(\partial_iA_j\left(\partial_iA_j-\partial_jA_i\right)
 +\left(\partial_iA_i\right)^2\right). 
\end{align} 
Notice that at this order the KK tensor field $\sigma_{ij}$ starts to play an 
important role in the interaction, and that all these vertices contain at most a single time 
derivative.

Let us proceed then to consider the point-particle action of a spinning particle 
\cite{Levi:2018nxp}. Since we are considering the spin-orbit sector, which is 
linear in the spins of the particles, it is sufficient here to take into account only 
the minimal coupling part of the effective action of each of the spinning particles. 
This part of the action reads \cite{Levi:2015msa}:
\begin{align} \label{mcact}
S_{\text{pp}}(\lambda)=&\int 
d\lambda\left[-m \sqrt{u^2}-\frac{1}{2} \hat{S}_{\mu\nu} \hat{\Omega}^{\mu\nu}
	-\frac{\hat{S}^{\mu\nu} p_{\nu}}{p^2} \frac{D p_{\mu}}{D \lambda}\right],
\end{align}
where $m$ is the mass, $u^{\mu}$ is the $4$-velocity, $p_{\mu}$ is the conjugate linear 
momentum, and $\hat{\Omega}^{\mu\nu}$ and $\hat{S}_{\mu\nu}$, are the generic angular 
velocity and spin variables of the particle, respectively. This form of the action in
eq.~\eqref{mcact} is covariant, as well as invariant under gauge of the rotational 
variables \cite{Levi:2015msa}. This is in contrast to the action presented in 
\cite{Hanson:1974qy,Bailey:1975fe,Porto:2005ac}, which does not have generic rotational 
variables, and does not include the last term in eq.~\eqref{mcact}. Note that both the mass 
and spin couplings play important roles in the spin-orbit interaction.

Additional worldline mass couplings are also required at N$^3$LO. In particular, we have the 
following two new Feynman rules for four-graviton couplings: 
\begin{align}
\label{eq:mphi^4}  \parbox{12mm}{\includegraphics[scale=0.6]{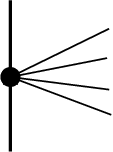}}
 = & -\frac{1}{24} m \int dt~\phi^4 ,\\
\label{eq:mphi^3A}  \parbox{12mm}{\includegraphics[scale=0.6]{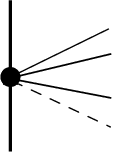}}
 = & \, \frac{1}{6} m \int dt~\phi^3A_iv^i,
\end{align}
where the thick vertical lines represent worldlines, and the spherical blobs 
represent mass insertions. 

Finally, let us consider the new worldline spin couplings required in this sector. 
For the two-graviton coupling to the worldline spin, the new Feynman rule is
\begin{align}
\label{eq:sphi^2}  \parbox{12mm}{\includegraphics[scale=0.6]{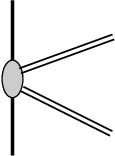}}
 = & \,\frac{1}{8} \int dt \big[S_{ij}\sigma_{il}
\left[2v^m\left(\partial_j\sigma_{lm}-\partial_l\sigma_{jm}\right)
+v^m\partial_m\sigma_{jl}+\partial_t\sigma_{jl}\right]\big],
\end{align}
where the oval blobs stand for the spin dipole sources.
Notice in particular the last term, which involves a time derivative, that enters here at the 
LO of the vertex; This did not occur in vertices at lower orders. These rules are already 
given in terms of the physical spatial components of the local spin tensor in the canonical 
gauge \cite{Levi:2015msa}, so all indices are Euclidean.
For the three-graviton coupling to the worldline spin, the new Feynman rule is
\begin{align}
\label{eq:sphiAsigma} \parbox{12mm}{\includegraphics[scale=0.6]{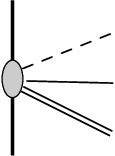}}
 = & \,\frac{c_d}{4} \int dt \left[S_{ij}\sigma_{il}
 \left(\partial_jA_l-\partial_lA_j\right)\phi\right],
\end{align}
and for the four-graviton coupling to the worldline spin, the new Feynman rule is
\begin{align}
\label{eq:sphi^3A}  \parbox{12mm}{\includegraphics[scale=0.6]{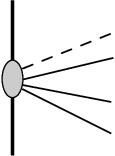}}
 = & \,\frac{c_d^3}{12} \int dt 
 \left[S_{ij}\partial_iA_j\, \phi^3\right].
\end{align}
Note that similar to what happens as of the NLO, at the N$^n$LO, the $(n{+}1)$-scalar 
graviton-spin coupling is absent in the KK fields together with our gauge choice for the 
rotational variables. Thus, this vertex is deferred to higher PN orders.

\section{Diagrammatic expansion} 
\label{Feynman}

In this section we present and evaluate the Feynman diagrams that 
comprise the perturbative PN expansion of the N$^3$LO spin-orbit 
sector at order $G^4$. As illustrated in table 
\ref{stateoftheart}, the analysis of this sector builds on the 
N$^2$LO spin-orbit sector obtained in 
\cite{Hartung:2011te,Marsat:2012fn,Bohe:2012mr}, and in 
\cite{Levi:2015uxa} via the EFT of spinning gravitating objects, 
and on the non-spinning N$^3$LO sector (at the $3$PN order) 
\cite{Jaranowski:1997ky,Jaranowski:1999ye,Blanchet:2000nv,
Blanchet:2000ub,Damour:2001bu,
Itoh:2003fy,Blanchet:2003gy,Foffa:2011ub,Levi:2011up}. In 
contrast to the non-spinning case, all possible topologies are 
realized in the spinning sector at each order of $G$ 
\cite{Levi:2008nh}. Hence, in the present sector
three-loop topologies (in the worldline picture, as specified 
below) must be tackled, including topologies whose integrals need 
to be reduced using integration by parts (IBP) 
\cite{Smirnov:2006ry}. This is unlike the situation in the 
non-spinning sector, where such three-loop topologies appear only 
at N$^4$LO (at the 4PN order) 
\cite{Levi:2018nxp}. 

We start by giving a comprehensive account of the 
topologies and the corresponding integrals that appear in this 
sector, before proceeding to enumerate all of the Feynman graphs 
and their evaluation. Similar to the ingredients presented in the 
previous section, all of the computational aspects of the 
work presented in this section were carried out using the 
\texttt{EFTofPNG} code \cite{Levi:2017kzq,Levi:2018stw}, that has 
been extended to handle this challenging sector.

\subsection{Topologies}
\label{topologies}

\begin{figure}[t]
\begin{minipage}{0.48\textwidth}
\centering
\includegraphics[width=0.1\linewidth]{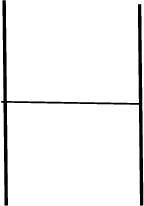}
\caption{The single graph topology at order $G$: One-graviton exchange with 
no self-interaction.}
\label{G1topo} 
\end{minipage}
\hspace{0.4cm}
\begin{minipage}{0.48\textwidth}
\centering
\includegraphics[width=0.35\textwidth]{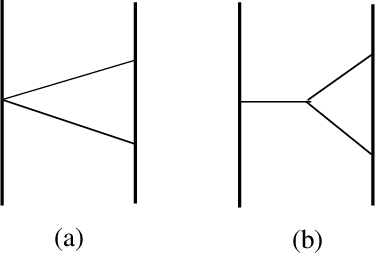}
\caption{Graph topologies at order $G^2$: 
(a) No self-interaction vertices. 
(b) One cubic vertex. This is a one-loop 
topology in the worldline picture.} 
\label{G2topo}
\end{minipage}
\end{figure}

\begin{figure}[t]
\begin{center}
\includegraphics[width=0.35\textwidth]{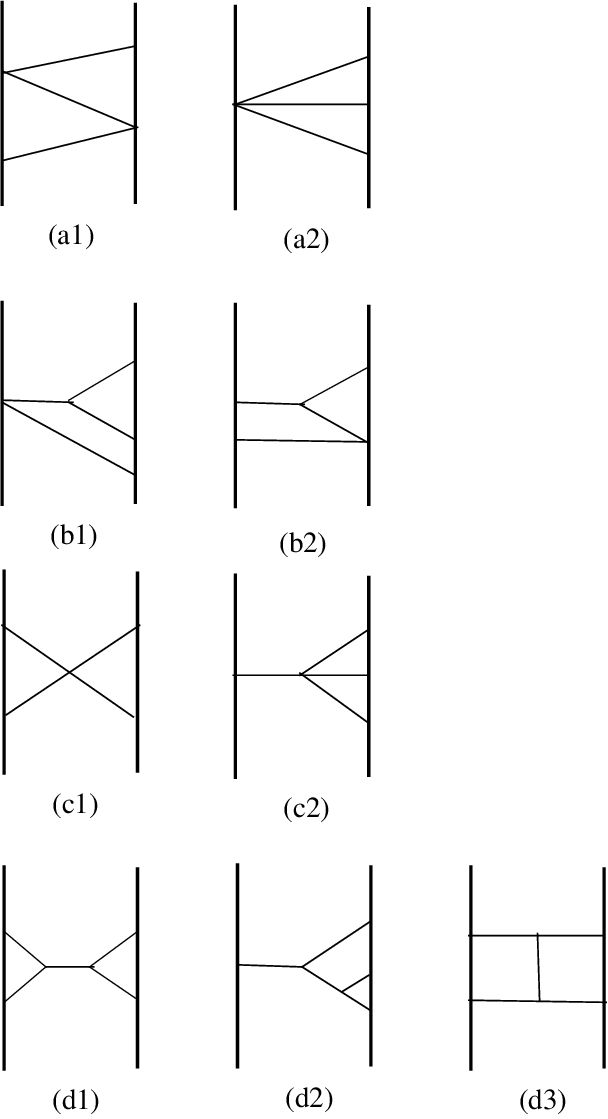}
\caption{Graph topologies at order $G^3$: 
(a) No self-interaction vertices. 
(b) One cubic vertex. 
(c) One quartic vertex. 
(d) Two cubic vertices.
Topologies (c) and (d) constitute two-loop topologies
in the worldline picture. 
The topology (d3) is a rank-two topology as specified below.} 
\label{G3topo}
\end{center}
\end{figure}

Let us begin by describing the generic topologies that enter at 
order $G^4$. To establish our terminology, we first review the 
topologies that appear at lower orders of $G$; these are shown 
for $G^1$, $G^2$, and $G^3$ in figures \ref{G1topo}, 
\ref{G2topo}, \ref{G3topo}, respectively \cite{Levi:2018nxp}.

First we consider the integral expressions which correspond to 
the various topologies. For instance, graph (a) in figure 
\ref{G2topo} is proportional to
\be \label{factorizable}
\text{Fig.~2(a)} \sim 
\int_{\vec{p}_1} \frac{e^{i\vec{p}_1\cdot\left(\vec{x}_1 - 
\vec{x}_2\right)}}{{\vec{p}_1}^2} 
\int_{\vec{p}_2} \frac{e^{i\vec{p}_2\cdot\left(\vec{x}_1 - 
\vec{x}_2\right)}}{{\vec{p}_2}^2},
\ee
so this is just a factorization into two copies of the basic 
topology at order $G$ shown in figure \ref{G1topo}. However, if 
we perform the following simple change of variables:
\be \label{changevar}
p_1+p_2\to p, \quad p_2 \to k_1,
\ee
we obtain
\be \label{standardize}
\text{Fig.~2(a)} \to 
\int_{\vec{p}} e^{i\vec{p}\cdot\left(\vec{x}_1 - \vec{x}_2\right)}
\int_{\vec{k}_1} \frac{1}{{\vec{k}_1}^2{(\vec{p}-\vec{k}_1)^2}}
\sim\text{Fig.~2(b)},
\ee
where $p$ is the Fourier momentum (or the momentum transfer of the source), and $k_1$ is the 
loop momentum. Hence, we see that both topologies at order $G^2$ can also be expressed in 
terms of a single basic one-loop integral. On the other hand, from eq.~\eqref{factorizable} 
which is written directly in the worldline picture, we see that the topologies that appear at 
each higher order in $G$ are often no more complicated than those appearing at lower orders.
It is thus beneficial to discern the worldline picture, in which all topologies and graphs in 
this paper are drawn. This is facilitated by considering the following useful definition of 
loop order in the worldline picture, which we will use to classify all graphs in this paper:

\begin{definition}
The loop order, $n_L$, of a graph at order $G^{n}$ in the 
worldline picture can be defined as
\be 
n_L\equiv 2n-\sum_{i=1}^{n+1} m_i,
\ee
where $m_i$ denote the numbers of gravitons in each of the $n+1$ 
worldline insertions.
\end{definition}
In particular, we see that the maximal loop order in the worldline 
picture at order $G^{n}$, which is $n-1$, is realized only in the topologies which contain 
exclusively one-graviton worldline insertions. For instance, figure \ref{G2topo}(b) is a 
one-loop topology, whereas the factorizable figure \ref{G2topo}(a) is a $0$-loop topology.

It is straightforward to generalize the change of variables in eq.~\eqref{changevar}, as well 
as to apply the definition above to topologies at higher orders in $G$. 
This gives us two possible perspectives on any given graph at order $G^{n+1}$: either in a 
form standardized into a $n$-loop integral, using a change of variables as in 
eq.~\eqref{changevar}, which corresponds to the quantum picture of a two-point function with 
massless propagators \cite{Kol:2013ega,Levi:2018nxp}; or in the worldline picture, with 
integrals that may factorize into a product with copies of the basic $0$-loop topology at 
order $G$. In particular, graphs with $m$-graviton worldline insertions, where $m\geq 2$, are 
always factorized in this way and correspond to lower-loop graphs in the worldline picture.

With these two perspectives in mind, the topologies at order $G^3$ (shown in figure 
\ref{G3topo}) can be classified into three types according to their two-loop standardized 
integral form, as detailed in \cite{Levi:2011eq}:
\begin{enumerate}
\item The \textit{nested} type, which includes most of the topologies at this order, 
\{(a1), (a2), (b1), (b2), (c2), (d2)\}.
\item The \textit{factorizable} type, which includes only the two topologies \{(c1), (d1)\}.
\item The topologies that are neither of the two former types, which must be reduced to a 
linear combination of the two former basic integrals using IBP relations. Only topology (d3) 
falls into this class. 
\end{enumerate}
In view of the above classification it is useful to introduce the following definition:
\begin{definition}
We define a topology at order $G^{n+1}$ to be of rank $r$, when $r$ of the basic $n$-loop 
integral types are required in order to express its $n$-loop integral form. 
\end{definition}
For instance, topology (d3) is a rank-two topology since it is expressed in terms of linear 
combinations of (the integrals corresponding to) nested and factorizable topologies. All 
other topologies at this order are rank-one topologies, since they are themselves nested or 
factorizable. 

\begin{figure}[t]
\includegraphics[width=0.74\textwidth]{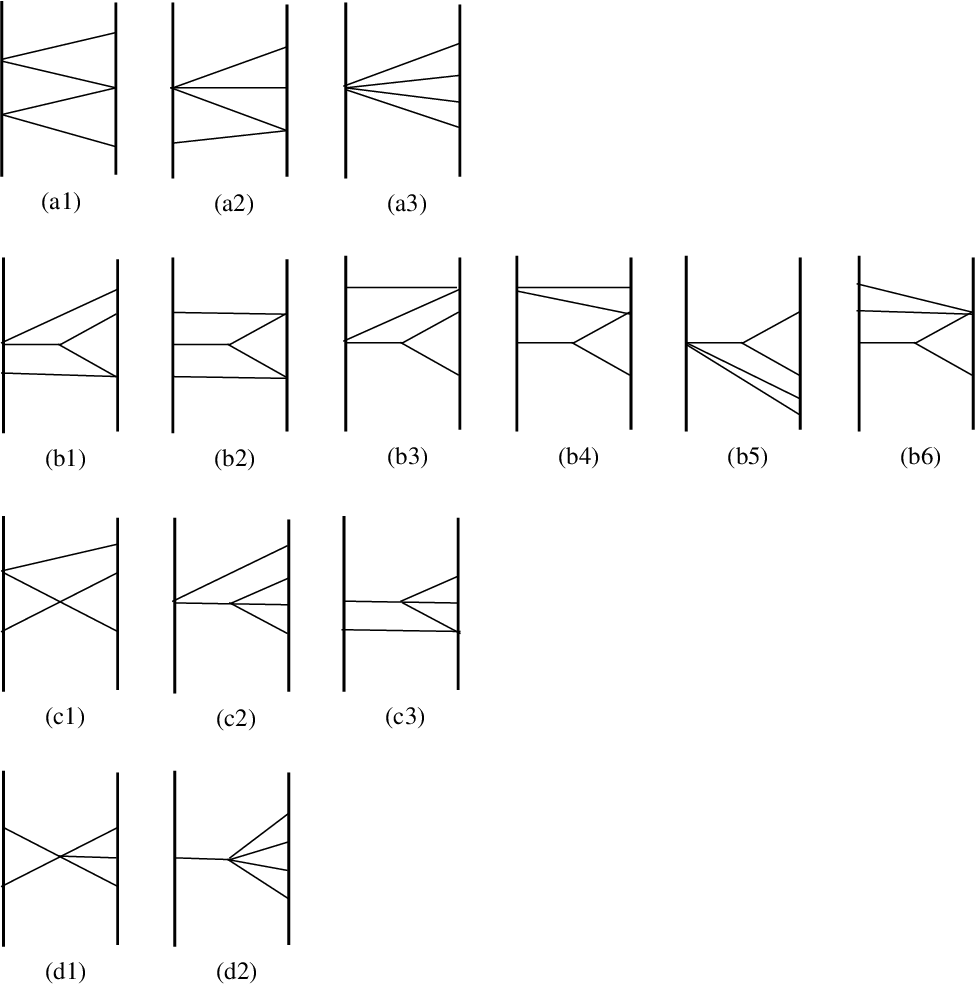}
\includegraphics[width=\textwidth]{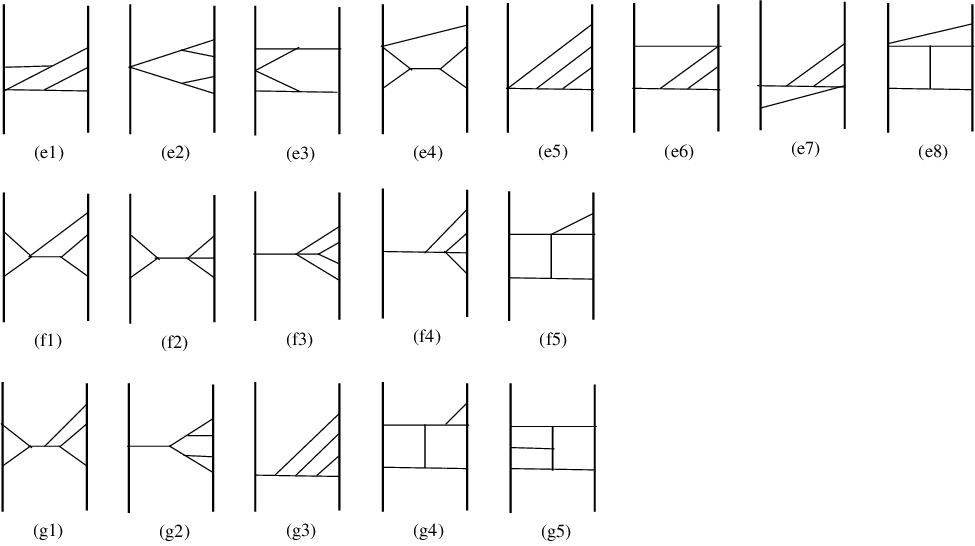}
\caption{Graph topologies at order $G^4$: 
(a) No self-interaction vertices. 
(b) One cubic vertex. 
(c) One quartic vertex. 
(d) One quintic vertex.
(e) Two cubic vertices. 
(f) One cubic vertex and one quartic vertex. 
(g) Three cubic vertices.}
\label{G4topo}
\end{figure}
Let us finally proceed to the topologies that appear in the spin-orbit sector at order $G^4$, 
which are shown in figure \ref{G4topo}. There are in fact three basic types of integrals, 
which form a basis for the topologies at this order. The $32$ topologies at this order 
can be classified as follows:
\begin{enumerate}
\item The \textit{nested-nested} type, which accounts for 22 of the topologies, 
\{(a), (b), (c2)-(c3), (d2), (e1)-(e3), (e5)-(e7), (f3)-(f4), (g2)-(g3)\}.
\item The \textit{factorizable-nested} type, which includes 
topologies \{(d1), (f1), (f2), (g1)\}. 
\item The \textit{nested-factorizable} type, which includes 
only the topologies \{(c1), (e4)\}.
This type cannot be realized as a three-loop topology in the worldline picture, where it 
actually contains only two-loop topologies. Any attempt to draw such a three-loop graph would 
contain graviton loops, which are purely quantum, and excluded in our setup. Of course, this 
type is still three-loop in the quantum two-point function picture \cite{Foffa:2019rdf}.
\item 
Beyond these basic types, there are higher-rank topologies, which can be further subdivided:
\begin{enumerate}
\item There is a single rank-two type, which can be expressed as a combination of the 
nested-nested and the nested-factorizable topologies. There is a single topology of this 
type, (e8). 
\item The rank-three topologies, \{(f5), (g4), (g5)\}, which can only be expressed in terms 
of linear combinations of all three basic integral types at this order.
\end{enumerate}
\end{enumerate}
The rank-one topologies all boil down to one-loop computations, whereas 
higher-rank topologies need to be worked out more laboriously.
Notice that only $12$ topologies at order $G^4$ are three-loop topologies in the worldline 
picture, namely topologies (d), (f), and (g) in figure \ref{G4topo} (see also figure 12 in 
\cite{Levi:2018nxp}). It is Feynman graphs with these topologies that are counted in the 
relevant entry in table \ref{stateoftheart}. As we shall see below in section \ref{findings}, 
these are also the graphs that give rise to the novel features that appear in this sector.

\subsection{Graphs}
\label{graphs}

We are now ready to enumerate the full set of Feynman graphs that contribute to the N$^3$LO 
spin-orbit sector at $G^4$. Clearly, as can also be understood from glancing at table 
\ref{stateoftheart}, the construction of the current sector builds on the N$^2$LO spin-orbit 
and the non-spinning N$^3$LO sectors, with the notable difference that the latter does not 
introduce three-loop graphs in the worldline picture \cite{Levi:2011up,Foffa:2011ub}. This is 
yet another sense in which the spinning sector is more complicated than the non-spinning 
sector, because all possible topologies are realized at each order in $G$, even when the KK 
field decomposition is used \cite{Kol:2007bc,Levi:2008nh,Kol:2010ze}. Further, the presence 
of a single spin coupling among the worldline insertions means that fewer graphs are 
equivalent under the permutation of worldline insertions, leading to more unique graphs in 
the spin-orbit sector than in the non-spinning sectors. In fact, there are far more distinct 
graphs in the N$^3$LO spin-orbit sector at $G^4$ than even at N$^5$LO in the non-spinning 
sector at $G^6$ \cite{Foffa:2019hrb,Blumlein:2019zku}. 

The generation of the Feynman graphs was carried out using the \texttt{FeynGen} module of the 
\texttt{EFTofPNG} public code \cite{Levi:2017kzq}, which was extended to this order. The 
extension of the code and the graphs were crosschecked. The graphs are drawn (using JaxoDraw 
\cite{Binosi:2003yf,Binosi:2008ig} based on \cite{Vermaseren:1994je}) in figures 
\ref{below3loopSOa-b4}--\ref{3loopSOg4-5} below. 
The higher-rank topologies are the more complex, and as such they give rise to the highest 
numbers of unique graphs. In general, the more self-interaction vertices in a topology, the 
more graphs it contributes to the sector. All in all, there are $388$ unique graphs in the 
sector; $174$ of these are three-loop graphs, and are shown in figures 
\ref{3loopSOd+f}--\ref{3loopSOg4-5}. 
$93$ of these are higher-rank graphs---in particular, $26$ are rank-two, and $67$ are 
rank-three. The higher-rank graphs are reduced using IBP methods, as discussed below. For 
comparison, in lower-order sectors only $17$ rank-two graphs contribute, yet these graphs 
still constituted the bottleneck of the calculation, even when automated using the first 
version of the public \texttt{EFTofPNG} code. 
Therefore, it was clear that more advanced integration methods, in particular for the 
reduction of higher-rank integrals, would be necessary.

\begin{figure}[t]
\centering
\includegraphics[angle=90,height=0.9\textheight,width=\textwidth]{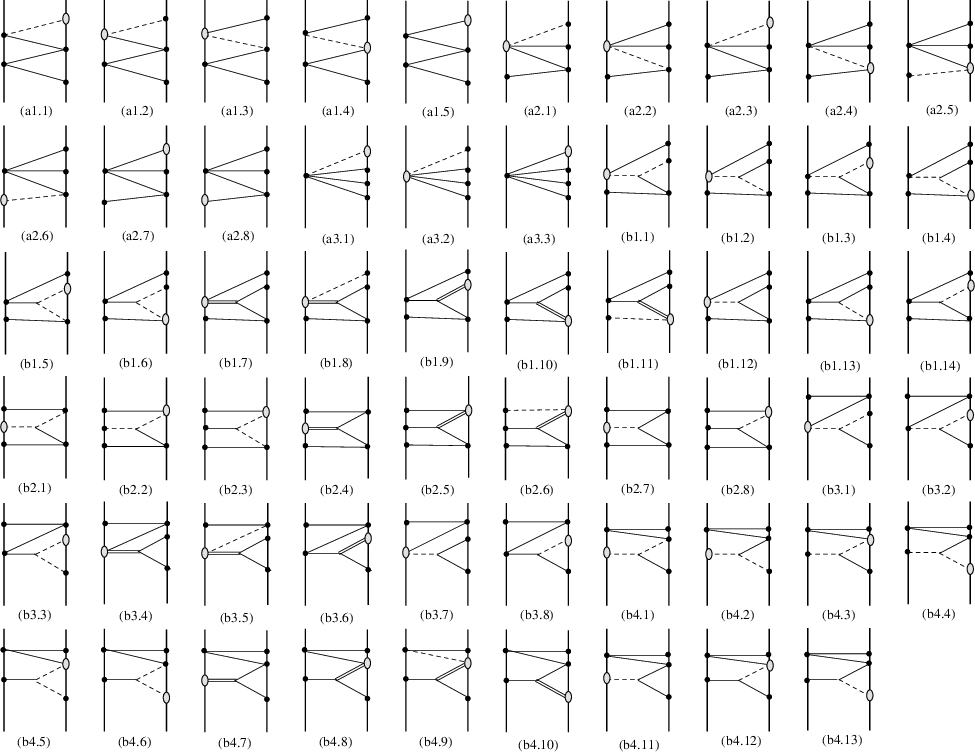}
\caption{Feynman graphs which contribute to the N$^3$LO spin-orbit coupling at $G^4$.
This figure contains all graphs with topologies (a) and (b1)-(b4) of figure \ref{G4topo}.
All of the graphs presented in this figure and the following ones should be accompanied by 
their `mirror' graphs, in which the worldline labels are exchanged, i.e.~$1\leftrightarrow2$.}
\label{below3loopSOa-b4} 
\end{figure}
\begin{figure}[t]
\centering
\includegraphics[angle=90,height=0.9\textheight]{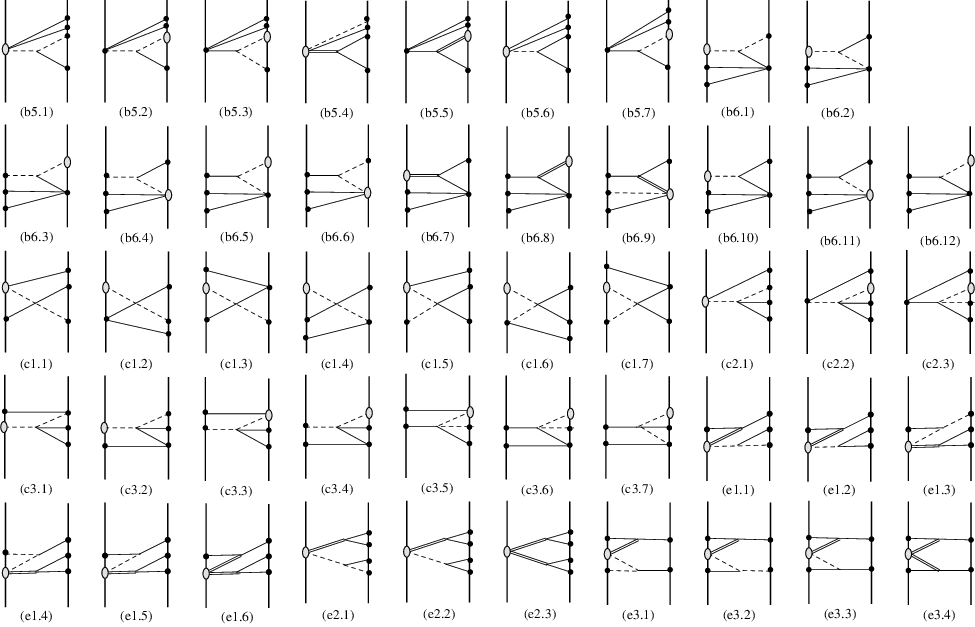}
\caption{Feynman graphs which contribute to the N$^3$LO spin-orbit coupling at $G^4$.
This figure contains all graphs with topologies (b5), (b6), (c), and (e1)-(e3), 
of figure \ref{G4topo}.}
\label{below3loopSOb5-6ce1-3} 
\end{figure}
\begin{figure}[t]
\centering
\includegraphics[angle=90,height=0.9\textheight]{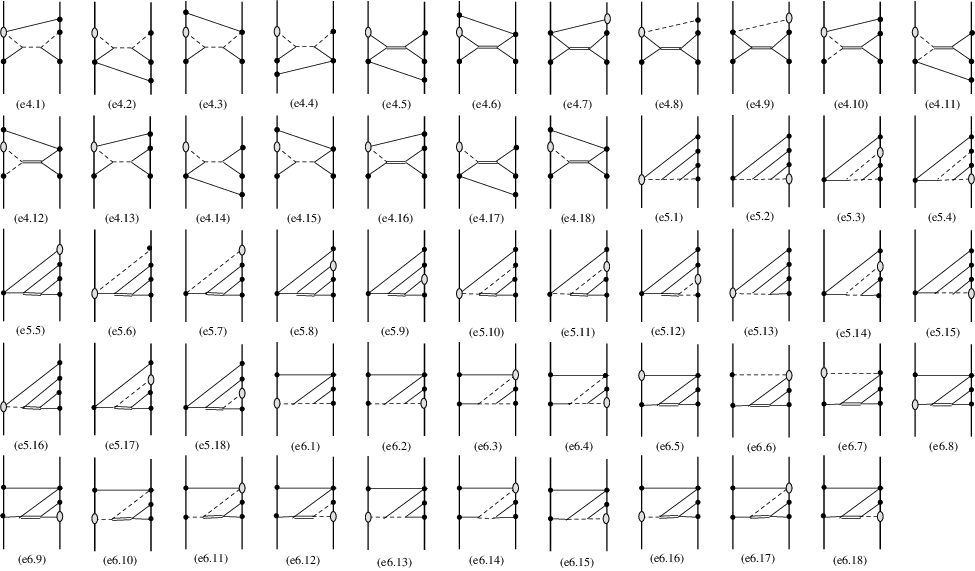}
\caption{Feynman graphs which contribute to the N$^3$LO spin-orbit coupling at $G^4$.
This figure contains all graphs with topologies (e4)-(e6) of figure \ref{G4topo}.}
\label{below3loopSOe4-6} 
\end{figure}
\begin{figure}[h!]
\centering
\includegraphics[angle=90,height=0.9\textheight]{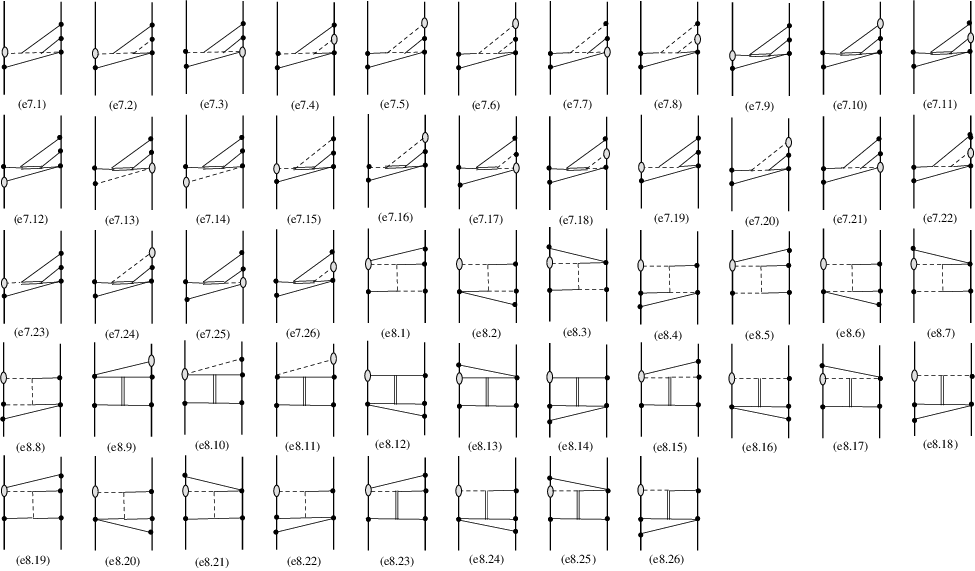}
\caption{Feynman graphs which contribute to the N$^3$LO spin-orbit coupling at $G^4$.
This figure contains all graphs with topologies (e7) and (e8) of figure \ref{G4topo}.
The graphs in group (e8) are rank-two graphs, which clearly factorize into the single 
rank-two topology (d3) in figure \ref{G3topo}.}
\label{below3loopSOe7-8} 
\end{figure}
\begin{figure}[h!]
\centering
\includegraphics[angle=90,height=0.9\textheight]{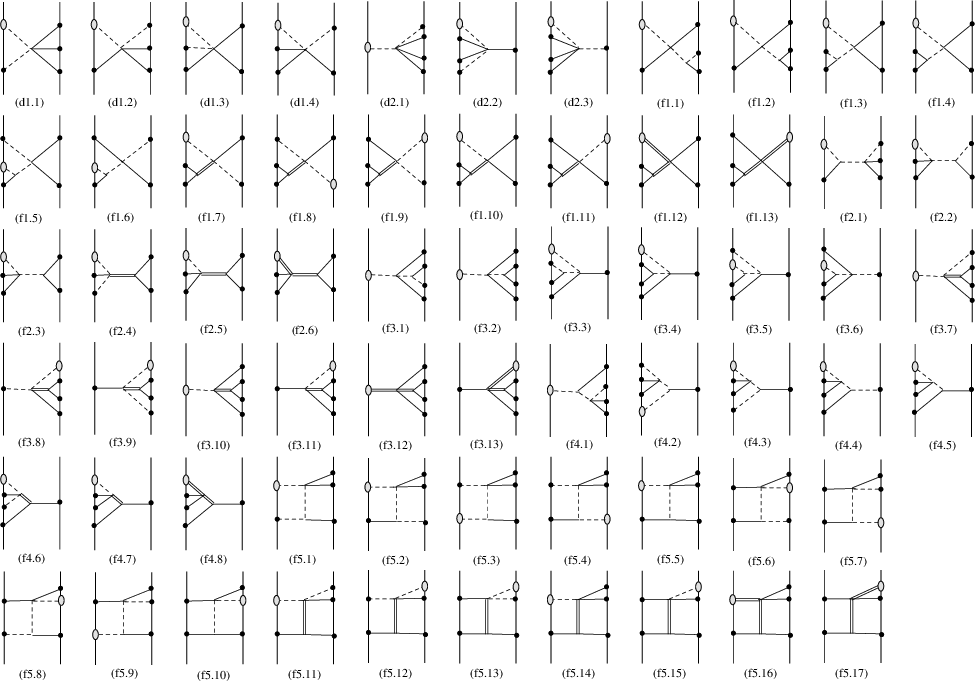}
\caption{Feynman graphs which contribute to the N$^3$LO spin-orbit coupling at $G^4$.
This figure contains all graphs with topologies (d) and (f) of figure \ref{G4topo}. The 
graphs in group (f5) are rank-three graphs that require reduction, and appear in the 
non-spinning sector only at N$^4$LO.}
\label{3loopSOd+f} 
\end{figure}
\begin{figure}[h!]
\centering
\includegraphics[angle=90,height=0.9\textheight,width=\textwidth]{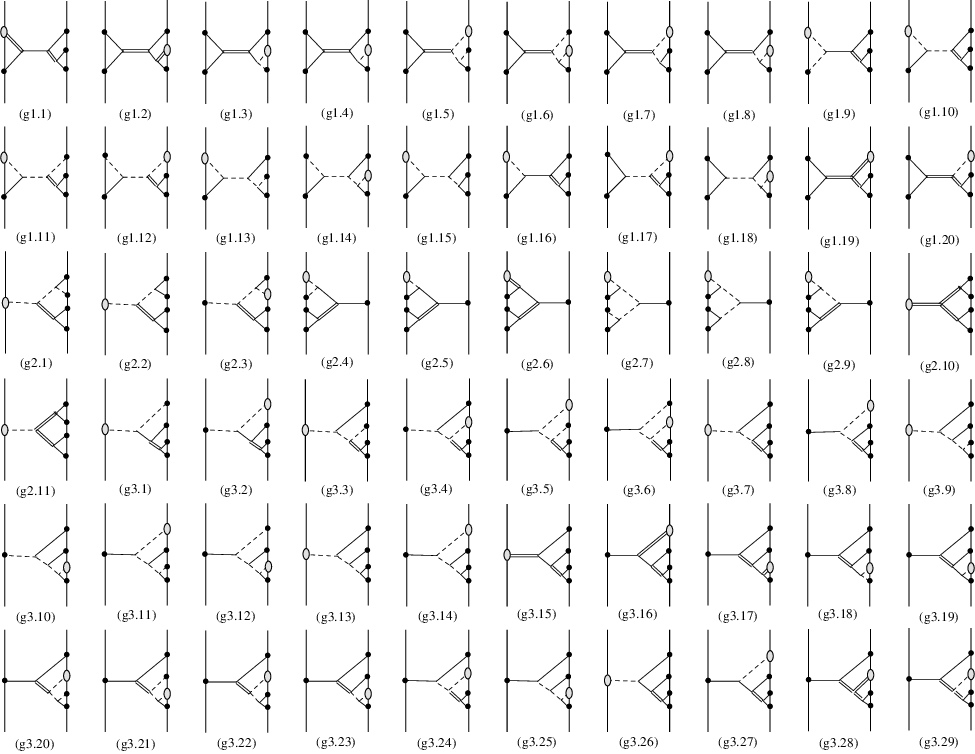}
\caption{Feynman graphs which contribute to the N$^3$LO spin-orbit coupling at $G^4$.
This figure contains all graphs with topologies (g1)-(g3) of figure \ref{G4topo}.}
\label{3loopSOg1-3} 
\end{figure}
\begin{figure}[h!]
\centering
\includegraphics[angle=90,height=0.9\textheight,width=\textwidth]{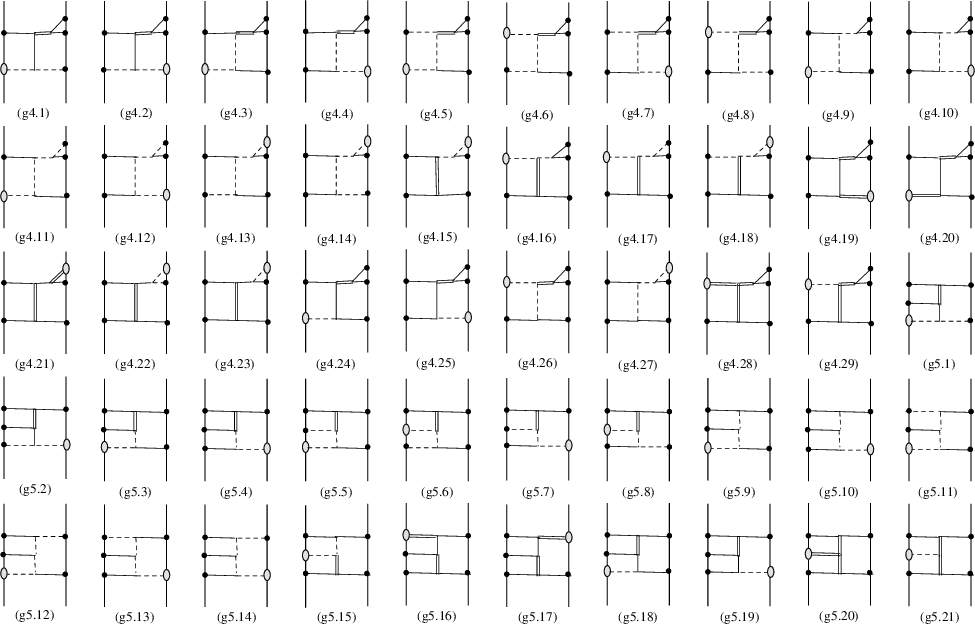}
\caption{Feynman graphs which contribute to the N$^3$LO spin-orbit coupling at $G^4$.
This figure contains all graphs with topologies (g4)-(g5) of figure \ref{G4topo}. These are 
rank-three topologies that are deferred in the non-spinning sector to the N$^4$LO.}
\label{3loopSOg4-5} 
\end{figure}

\subsection{Integration and scalability}
\label{scalability}

To evaluate the graphs in this sector we made use of the public \texttt{EFTofPNG} code, with 
significant upgrades to its \texttt{NLoop} and \texttt{Main} modules \cite{Levi:2017kzq}. 

Higher-rank integrals are reduced using the integration by parts (IBP) method 
\cite{Smirnov:2006ry}, which was previously implemented within the \texttt{EFTofPNG} code for 
rank-two integrals, with the IBP reduction done `by hand'. IBP relations were first applied 
to two-point graphs with three-loop massless propagators in \cite{Chetyrkin:1981qh}, and the 
algorithmic reduction was developed in \cite{Laporta:1996mq,Laporta:2001dd}. The main 
development in the \texttt{EFTofPNG} code involves a new implementation of the algorithmic 
IBP method.

The upgraded code processes the integrals in this sector in several stages. First, the 
integrations are streamlined by separating into an external Fourier momentum and three 
internal loop momenta, similarly to the change of variables used in eqs.~\eqref{changevar} 
and \eqref{standardize}. Next, all tensor dependence in the loop momenta is removed from the 
numerators via the projection method \cite{Karplus:1950zza,Kniehl:1990iva,Binoth:2002xg}, 
which may be familiar to the readers from the Passarino-Veltman procedure at one-loop 
\cite{Passarino:1978jh} (see also \cite{Boels:2018nrr,Chen:2019wyb} for a useful modern 
presentation of the method). Since spins are derivatively coupled, tensor numerators are 
present as high as rank eight, similar to the rank encountered at the N$^5$LO in the 
non-spinning sector. Finally, the resulting integrals are reduced using IBP relations 
\cite{Smirnov:2006ry} via a variant of Laporta's algorithm \cite{Laporta:2001dd}, resulting 
in expressions in terms of only three basic scalar integrals, corresponding to the three 
basic topology types at order $G^4$, specified in section \ref{topologies}. Altogether, the 
code runs over approximately two days.

This upgrade will be made publicly available in a future update of the code, and will be 
presented in a forthcoming publication.

\subsection{Findings}
\label{findings}

The values of the individual Feynman graphs are found in the ancillary files to this 
publication (both in PDF and in machine-readable files). When two graphs are related by 
exchange of worldline labels, $1\leftrightarrow 2$, we present only the graph with spin 
coupling on worldline ``1''. All the new results in this work were confirmed via 2 independent 
implementations of the formulation and the machinery presented in the previous sections.

Let us point out here some notable features of our results. Of these features, those which 
are new to the spinning sector arise uniquely from three-loop topologies and thus did not 
show up at the N$^3$LO non-spinning sector at $G^4$ within EFT derivations 
\cite{Levi:2011up,Foffa:2011ub,Blumlein:2019zku}.

\paragraph{Zeros.}
There are a total of 40 graphs that vanish in the present sector.
Of these, 25 can be understood due to the presence of contact interaction terms.
The nested-factorizable topologies (c1) and (e4) in figure \ref{G4topo} yield zeros 
due to their dependence on $\epsilon\equiv d-3$. Such zeros appear at order $G^3$, where the 
factorizable two-loop topologies behave as:
\be
\text{Fig.~3(c1),(d1)}\propto [\Gamma(-\epsilon)]^{-1} \sim \epsilon + {\cal{O}}(\epsilon^2).
\ee
At order $G^4$ the nested-factorizable topologies (c1) and (e4) in figure \ref{G4topo} contain 
the factorizable two-loop topologies as their subgraphs, and we then have for them: 
\be \label{nestedfactor}
\text{Fig.~4(c1),(e4)}\propto \zeta(2)[\Gamma(-\epsilon)]^{-1}  
\sim \zeta(2)\epsilon + {\cal{O}}(\epsilon^2),
\ee
which gives rise to the appearance of $\zeta(2)=\pi^2/6$ in rank-three graphs
as we discuss below. To recap, graphs of the nested-factorizable 
topologies, similar to the factorizable two-loop topologies, stand here for purely 
short-distance contributions, which are contact interaction terms of the form 
$\delta(\vec{r})$, with $\vec{r}\equiv\vec{r}_1-\vec{r}_2$ \cite{Levi:2011eq}. 

The remaining $15$ graphs are sporadic and vanish due to a variety of miscellaneous reasons. 
%8 before d->3? 

\paragraph{Riemann zeta values.}
The rank-three topologies (f5), (g4), and (g5), give rise to terms proportional to 
$\zeta(2)=\pi^2/6$, and are the source of all transcendental contributions to our result, the 
first such observed in the spinning sector. We recall that the rank-three topologies involve 
a linear combination of all three basic three-loop integrals, and that these $\zeta(2)$ 
factors in particular originate uniquely from the nested-factorizable integral as noted in 
eq.~\eqref{nestedfactor} above. Now, we recall that the IBP reduction relations yield linear 
combinations of basic integrals, with the coefficients including also the factor 
$\epsilon^{-1}$, i.e.~with explicit poles in $\epsilon$, see e.g.~\cite{Levi:2011eq}. Hence, 
in these rank-three topologies we have poles canceling out the zeros in $\epsilon$, and the 
$\zeta(2)$ factor is then uncovered. There are $31$ graphs of the three rank-three topologies 
that give rise to terms that contain $\zeta(2)$.

Riemann zeta values occur in quantum loop corrections starting at one loop, and 
thus in view of the contact interaction terms, associated with encapsulated UV physics as 
noted above, that arise at the N$^2$LO, it is not surprising that such Riemann zeta values 
appear in the related graphs at N$^3$LO. 
Let us point out that topology (e8), which is a rank-two topology, comprised from two of the 
three basic integrals, also contains the nested-factorizable integral with the $\zeta(2)$ 
factor, yet it does not yield terms with $\zeta(2)$. This is because topology (e8) is 
trivially factorized into the rank-two topology (d3) in figure \ref{G3topo}, which appears at 
the N$^2$LO, where such transcendental numbers do not emerge; Stated differently, topology 
(e8) is simply a two-loop topology in the worldline picture.

\paragraph{Simple poles and logarithms.}
Most of the three-loop graphs in the worldline picture (corresponding to graphs with one of 
the topologies (d), (f), or (g) in figure \ref{G4topo}) yield simple poles in 
$\epsilon\equiv d-3$ in conjunction with logarithms in $r/R_0$. 
This is because these graphs contain contributions proportional to the factor
$\Gamma(\epsilon)(r/R_0)^{-4\epsilon}$, which gives rise to these poles and logarithms upon 
expansion in $\epsilon$. In this sector all of the aforementioned topologies (except the 
three nested-nested topologies (d2), (f3), and (g2)) yield such terms, namely both the 
factorizable-nested and the nested-nested basic types. All in all, $131$ of the three-loop 
graphs give rise to such terms.
%proportion of poles-logs $-4$

Let us highlight again that only the two basic types of factorizable-nested and nested-nested 
integrals, when they occur within the three-loop topologies in the worldline picture, give 
rise to poles in $\epsilon$ and logarithms, whereas only the nested-factorizable integrals, 
when contained within three-loop graphs, gives rise to the $\zeta(2)$ factors. Thus, these 
two behaviors occur entirely independently:
there are numerous rank-one three-loop graphs with poles and logarithms and no $\zeta(2)$ 
factors, and conversely the rank-three graph (g5.1) in figure \ref{3loopSOg4-5}, which yields 
a $\zeta(2)$ factor, does not involve any poles or logarithms.

\section{\texorpdfstring{N$^3$LO}{N3LO} gravitational spin-orbit action at 
\texorpdfstring{$G^4$}{G4}}
\label{result}

Adding up all of the graphs we get the following contribution to the N$^3$LO spin-orbit 
Lagrangian at $G^4$:
\begin{align}
L_{\text{SO}}^{\text{N$^3$LO}} & 
=\frac{G^4}{r^5} \frac{\vec{S}_1}{m_1}\cdot \vec{v}\times\vec{n} 
\Bigg[\Bigg(\frac{20}{3} - 13\,\zeta(2) 
- \frac{5}{3} \bigg(\frac{1}{\epsilon} - 4 \log\left(r/R_0\right)\bigg) 
\Bigg)\, m_1^3 m_2^2 -\frac{31}{3}\, m_1^2 m_2^3 \nn\\
& \qquad \qquad \qquad \qquad  
- \frac{3}{8}\, m_1 m_2^4 \Bigg] 
+ \big[ 1\leftrightarrow2 \big],
\end{align}
where $\epsilon_{ijk}S_k=S_{ij}$, with $\epsilon_{ijk}$ the 
$3$-dimensional Levi-Civita symbol,
$\vec{v}\equiv \vec{v}_1-\vec{v}_2$, and $\vec{n}\equiv \vec{r}/r$. 
Note the overall dependence only on the relative velocity, $\vec{v}$, rather than on each of 
the worldline velocities separately, similar to the total results from topologies at the 
highest order in $G$ at lower PN orders in the spin-orbit sector 
\cite{Levi:2010zu,Levi:2015uxa}. 

Notice that all the contributions proportional to $m_1^3 m_2$ conspire to cancel out from the 
total result, though they appear in many topologies and individual graphs' values. This 
vanishing coefficient would have been the leading contribution in the test-particle limit, 
where the extreme-mass-ratio limit holds, i.e.~$m_1\gg m_2$, and the body ``2'' can be 
considered as a test particle moving in the Kerr metric (in harmonic coordinates) generated by 
body ``1''.
%As to this observation - similar at NLO, but not at NNLO.

Interestingly, the poles, logarithms, and factors of $\zeta(2)$ eventually appear
only in the term proportional to $m_1^2m_2^2$. Note that the appearance of these
features in the total result differs from the situation in the non-spinning sector within EFT 
derivations, where all the poles in $\epsilon$, 
logarithms, and Riemann zeta values, which show up in general as of the N$^3$LO, conspire to 
cancel out in each of the N$^n$LO sectors at $G^{n+1}$ for $n\leq5$, so that they all 
contain only finite terms with only rational coefficients \cite{Blumlein:2019zku}.

\section{Conclusions}
\label{lafin}

In this paper we have computed for the first time the 
contribution to the N$^3$LO gravitational spin-orbit coupling 
from interaction at $G^4$ via the EFT of spinning gravitating 
objects \cite{Levi:2015msa}. The computation is carried out in 
terms of Feynman diagrams with topologies at this order (see 
figure \ref{G4topo}), and relied on extending and developing the 
\texttt{EFTofPNG} public code \cite{Levi:2017kzq}. 
This constitutes the most computationally challenging part of the 
N$^3$LO spin-orbit sector, as far as integration is concerned, 
because it contains all possible topologies at $G^4$, including 
three-loop level. This sector enters at the 4.5PN order for 
maximally-rotating compact objects, and completes the results 
of \cite{Levi:2019kgk} at this order, thus pushing the current 
state of the art to the 4.5PN accuracy. 

The N$^3$LO spin-orbit coupling at $G^4$ consists of $388$ 
distinct graphs to evaluate. Of these graphs $174$ are genuine
three-loop graphs in the worldline picture, which give rise to 
new special features in the spinning sector, that also show up in 
the final result of the topologies at this order. 
These features include simple poles and 
logarithms that arise from dimensional regularization, as well as
the appearance of transcendental factors, which can be understood 
as next-order corrections of purely short-distance contributions 
that vanish in the classical context. This is in contrast to the 
non-spinning sector within EFT computations, where such special 
features conspire to cancel out from the final result in each of 
the N$^n$LO sectors at $G^{n+1}$ for $n\leq5$. 
% We also find that this piece of the sector vanishes in the test-particle limit.

We provide here a comprehensive account of the topologies in the worldline picture of the EFT 
approach. Further, all the computational aspects of the work are carried out via the unique 
\texttt{EFTofPNG} code. Due to the increased intricacy of this sector we have developed the 
\texttt{EFTofPNG} code, incorporating further techniques from the realm of particle 
amplitudes. 
We expect these developments to be extremely useful for studies of PN gravity, 
and we plan to release a public update to the \texttt{EFTofPNG} code to be presented in a 
forthcoming publication. The present sector illustrates once again that not only is tackling 
spins in gravity conceptually challenging, it is also more computationally challenging, as 
the higher conceptual intricacy is also reflected at the computational level in various 
aspects of the calculations. 

To complete the N$^3$LO spin-orbit sector all of the remaining 
contributions from topologies at up to order $G^3$ should 
also be evaluated. Of the latter, the contribution at $G^3$ is 
the most computationally demanding as it 
is the only remaining one with higher-rank graphs that require 
reduction, on top of the large number 
of contributing graphs. Nonetheless, we believe that our extensions of the
\texttt{EFTofPNG} code are well-suited to handle this challenge. 
In general, lower orders in $G$ are challenging due to the 
higher-order spin couplings and the proliferation of time derivatives. 
These contributions from lower 
orders in $G$ will also fix the resolution of the simple pole and 
logarithm that are left in the total 
result here. Thus, the completion of the N$^3$LO spin-orbit at all 
orders in $G$ will be reported in forthcoming publications.

Finally, it is evident that such high-precision computations require 
crosschecks, preferably via independent PN methodologies. Therefore, 
prospective studies that overlap with these sectors, possibly incorporating 
further modern amplitudes methods as in 
\cite{Cachazo:2017jef,Guevara:2017csg,Cheung:2018wkq,Bern:2019nnu,Bern:2019crd}, 
are extremely desirable.

\acknowledgments

ML receives funding from the European Union's Horizon 2020 research and 
innovation programme under the Marie Sk{\l}odowska Curie grant agreements 
No.~847523 and No.~764850, and from the Carlsberg Foundation. 
ML has also been supported by the European Union's Horizon 2020 Framework 
Programme FP8/2014-2020 ``preQFT'' starting grant No.~639729. 
ML is grateful to Freddy Cachazo for the warm hospitality at Perimeter 
Institute where the final stages of this work were carried out.
AJM and MvH are both supported by an ERC starting grant No.~757978 and 
a grant from the Villum Fonden No.~15369. 
AJM is also supported by a Carlsberg Postdoctoral Fellowship (CF18-0641).
MvH is also supported by the European Union's Horizon 2020 research and 
innovation programme under grant agreement No.~793151.

\bibliographystyle{jhep}
\bibliography{gwbibtex}

\end{document}